\documentclass[journal,article,submit,moreauthors,pdftex]{Definitions/mdpi} 

\usepackage{epsfig}
\usepackage{subfigure}          
\usepackage{placeins}
\firstpage{1} 
\pubvolume{xx}
\issuenum{1}
\articlenumber{5}
\pubyear{2019}
\copyrightyear{2019}
\history{Received: date; Accepted: date; Published: date}

\pdfoutput=1

\Title{A Hybrid Model for Lift Response to Dynamic Actuation On A Stalled Airfoil}

\Author{Xuanhong An $^{1,}$*, David R. Williams $^{2}$ and Maziar S. Hemati $^{3}$}

\AuthorNames{Xuanhong An, David R. Williams and Maziar S. Hemati}

\address{%
$^{1}$ \quad Department of Mechanical \& Aerospace Engineering, Princeton University, Princeton, NJ 08544, USA; xuanhong@princeton.edu\\
$^{2}$ \quad Department of Mechanical, Materials, and Aerospace Engineering, Illinois Institute of Technoogy, Chicago, IL 60616, USA; williamsd@iit.edu\\
$^{3}$ \quad Department of Aerospace Engineering \& Mechanics, University of Minnesota, Minneapolis, MN 55455, USA; mhemati@umn.edu}

\corres{Correspondence: * xuanhong@princeton.edu}

\abstract{The current research focuses on modeling the lift response due to {dynamic (time-varying) `burst-type'} actuation on a stalled airfoil.
{Dynamic `burst-type' actuation exhibits two different characteristic dynamic behaviors within the system, namely the high-frequency and low-frequency components. These characteristics introduce modeling challenges.
In this paper, we propose a hybrid model composed of two individual sub-models, one for each of the two frequencies.}
  {The lift response due to high-frequency single burst actuation is captured using a convolution model.
    The low-frequency component due to nonlinear burst-burst interactions are captured using a Wiener model, consisting of linear time-invariant dynamics and a static output nonlinearity.
The hybrid model is validated using data from wind tunnel experiments.}
}


\usepackage{tikz}

\begin{document}
	\nolinenumbers


\section{Introduction}\label{subsec:intro}

Unsteady flow separation causes transient aerodynamic forces in a variety of fluid dynamic applications and leads to performance {degradation in many} devices. For example, the dynamic stall vortices formed on helicopter rotor blades contribute to the unbalanced lift force and result in a undesired roll moment. In another example, by removing the pilot, who is the major limitation of the high performance aircraft maneuverability, the next generation unmanned aircraft could achieve super maneuverability. Then the unsteady flow separation, which is an inherent phenomenon of the super maneuverability, may become the limiting factor to performance. Another common type of fluid dynamic application where the unsteady flow separation needs to be addressed is the vertical axis wind turbine. Because their blades constantly change angle of attack relative to the incoming flow at a fast rate, the torque that can be produced becomes limited by the unsteady flow separation. Therefore, alleviating or even eliminating the unsteady flow separation could benefit many fluid dynamic applications.  

Active flow control (AFC) has attracted substantial attention for decades partly due to its ability to reattach the separated flow on stalled airfoils. Investigations by Williams et al.\cite{williams2010unsteady}, An \cite{an2015feedforward} , Williams et al. \cite{williams2015dynamic} and Williams and King \cite{williams2018alleviating} have shown that with a properly designed controller, AFC is also an effective way of alleviating the unsteady {separated} flow which is always in the transient state. Under this circumstance, the "dynamic actuation" is needed. The definition of dynamic actuation is that the actuation is time-varying. The time-varying actuation leads to transient aerodynamic force response, that cannot be modeled by quasi-steady models. Therefore, in order to design a proper controller, a dynamic model of predicting the aerodynamic force (e.g., lift, drag) {response due to} dynamic actuation is desired.
{These models are known as ``plant models'' within the controls literature.}
%
An accurate plant model {can} benefit controller {design}. First, a feedback controller can be {with less conservatism} when the plant model is accurate and reliable. Second, a model based observer (e.g., Kalman filter) can be achieved to alleviate the noisy feedback measurement without introducing any delay into the system.   

Previous investigations by Darabi and Wygnanski \cite{darabi2004active} and Amitay and Glezer \cite{amitay2002controlled} showed that burst actuation with short duty cycles is the most effective way of reattaching the separated flow on a stalled airfoil. However, in order to achieve the maximum lift gain, the burst frequency is much higher than the characteristic frequency contained in the unsteady aerodynamic forces that need to be controlled. Thus, another lower-frequency signal is superposed on the high-frequency burst signal as amplitude modulation. In the rest of this paper, we will refer to the high-frequency burst signal as the ``carrier wave'', the low-frequency signal as the ``control signal'' and their combination as the ``actuation signal''. Thus, the model of predicting the aerodynamic force (eg. lift) in response to the amplitude modulated input signal (eg. input voltage or the actuation momentum coefficient to the actuators) is desired.  


The classic approach of modeling the dynamic actuation is to employ the governing equation of the fluid and develop physical based models. For example, {Darakananda} et al. \cite{darakjdeprf18} developed a vortex sheet-point vortex model that could be utilized for modeling the dynamic actuation. In order to simulate the actuation at the leading edge, the critical leading-edge suction parameter (LESP) is adjusted with time. Furthermore, by adding a ensemble Kalman filter, a good agreement between the model and
immersed boundary method (IBM)
{high-fidelity numerical} simulations was obtained. However, the high-dimension feature of this vortex sheet-point vortex model makes it hard to be implemented into real-time control applications.

Williams, et al. \cite{williams2010unsteady} developed a simple linear model to predict the lift force variation associated with time-varying (transient) leading-edge actuation on a semi-circular wing. This model is achieved by averaging a family of models identified from a series of pseudo-random binary signals with different amplitudes. The model works well when the system is running near its design point; however, when the system is running at a point further away from its design point, the performance {deteriorates}. Later on, An et al. \cite{an2016modeling} and Williams et al. \cite{williams2015dynamic} employed a similar modeling strategy to model the synthetic (zero-net-mass flux) actuation on a nominal 2-D wing. In this case, the strong 3-D effect is absent compared with the semi-circular wing, and the complexity of the system is reduced. However, even for this simplified system, the averaged linear model only works well near its design point. In fact, the investigation by An et al. \cite{an2016modeling} reported that the nonlinear static gain in the system is the major cause of the linear model failure when the model is linearized at any point on the static nonlinear gain map. On the other hand, this model is only capable of capturing the low-frequency component in the lift variation with respect to the control signal only, since it only uses the low-frequency control signal as the input to the system. From this point, it is natural to investigate predictive modeling based on the complete actuation signal with both the high-frequency carrier wave and the low-frequency control signal. 

Following this idea, Williams \cite{williams2010unsteady} introduced a convolution model to predict the lift variation utilizing the actuation signal.  In their convolution model, they use the lift response to a single burst (impulse) signal as the kernel of the convolution integral. They demonstrated that the convolution model is capable of capturing both the high-frequency and low-frequency components in the lift variation. However, further investigations by An et al. \cite{an2016modeling} discovered that the convolution model fails when the burst frequency is high. They recognized that when the bursts are close to each other, the nonlinear burst-burst interaction becomes stronger, which can not captured by a linear convolution model.

In this paper, we propose a novel strategy to model this type of actuated system. The model includes two components, the first one is a modified convolution model that captures the response to the high-frequency component of actuation. The second component is a low-order Wiener model \cite{thomson1955response}\cite{wills2013identification}, consisting of linear dynamics with static output nonlinearity that is capable of modeling the low-frequency component in the lift response associated with the low-frequency control signal. 

The next section \ref{subsec:exp} describes the experimental setup and explains the actuation signal in more detail. In section \ref{subsec:result}, some preliminary results for the dynamic actuation are shown. In section \ref{subsec:7conv} the convolution model is discussed and modified. In section \ref{subsec:HW}, a Wiener model is proposed to model the low-frequency component of the lift. In section \ref{subsec:7Hybrid} the hybrid model consisting of the modified convolution model and the Wiener model is proposed. The hybrid model is then validated in section \ref{subsec:Results}. The conclusions are given in section \ref{subsec:conclusion}. 

\section{Experimental setup}\label{subsec:exp}
\begin{figure}[h]
	\centering
    \includegraphics[width=0.8\textwidth]{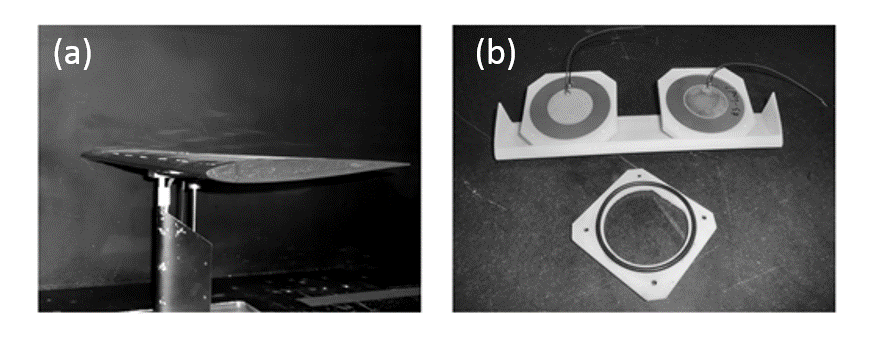}
    \caption{The NACA0009 wing and the actuators, (a) the wing inside the wind tunnel, (b) the actuators.} 
    \label{fig:wing} 
\end{figure}

The experiments were conducted in the Andrew Feier Unsteady Wind Tunnel at Illinois Institute of Technology. The test section of the wind tunnel is $2000 mm$ long, $600 mm$ wide and $600 mm$ high. The nominally two-dimensional NACA0009 wing with chord length $c=245 mm$ and wingspan $b=596 mm$ was used as the test article. The angle of attack was fixed at $20^o$. Eight piezoelectric (zero-net-mass-flux) actuators are installed at the leading edge of the wing, and the actuator orifices are located $0.05c$ from the leading edge on the suction side. The white strips in Figure \ref{fig:wing}a show the actuator orifices on the wing. The orifices have a $30^o$ angle relative to the local tangential direction. The eight piezoelectric actuators are grouped in four pairs for manufacturing simplification. Figure \ref{fig:wing}b shows the detail of one of the four pairs of actuators. The freestream velocity was fixed at $U_{\infty}3m/s$ throughout all the test cases in this paper, corresponding to a convective time $t_{convect}=\frac{c}{U_{\infty} = 0.082 s}$, where $t^+=\frac{t}{t_{convect}}$. The chord-based Reynolds number is Re=49000. The dimensionless excitation momentum coefficient $C_{\mu}=\frac{\rho U_j^2 A_{actuators}}{0.5 \rho U_{\infty}^2 A_{wing}}$, where $\rho$ is the air density, $U_j$ is the actuation jet speed in quiescent air, $A_{actuators}$ is the area of the actuators orifices and $A_{wing}$ is the wing area. The force and moment were measured by an ATI NANO-17 force transducer at $0.3c$. All the data sets were phase averaged utilizing the input signal to the actuators.   

\FloatBarrier

\begin{figure}[h]
	\centering
	\mbox{\subfigure[]{\epsfig{figure=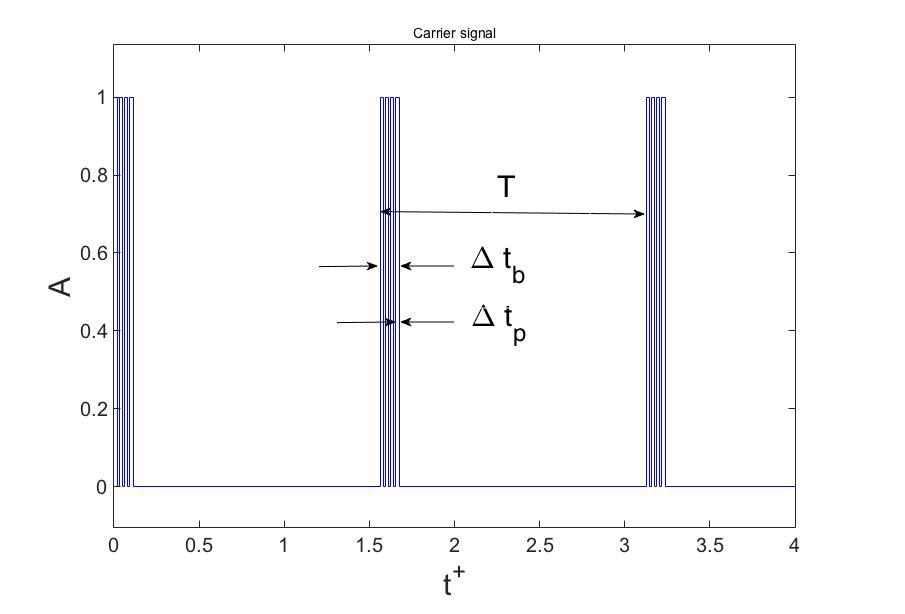, width=3.5in}}\quad} 
	
	\mbox{\subfigure[]{\epsfig{figure=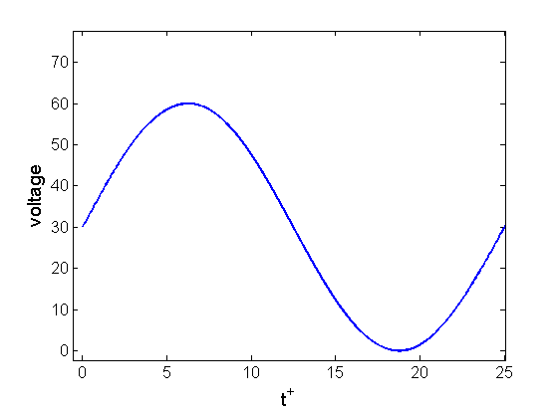, width=2.5in}}\quad} 
	~
   	\mbox{\subfigure[]{\epsfig{figure=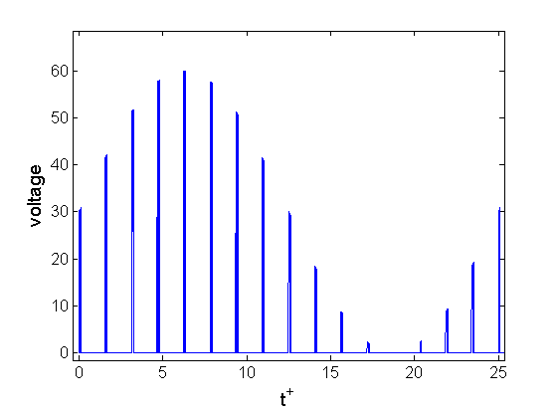,width=2.5in}}\quad}

    \caption{The actuator input signals, (a) the carrier wave, (b) the low-frequency control signal(c) the combined actuator input signal.}
    \label{fig:AFC_in}		
\end{figure}

The actuation input signal is modulated by three signals associated with three characteristic frequencies. The first is a pulse signal set to be at the resonant frequency of the actuators, $400 Hz$ with a pulse width of $\Delta t_p=0.0187t^+$. The second signal has a frequency $8 Hz$ corresponding to the period of $T=1.56t^+$ and a duty cycle of $\Delta t_b=0.125t^+$. The first two signals, namely the carrier wave and pulse signal, are exhibited in Figure \ref{fig:AFC_in}a. This frequency is used for maximum lift increment. The last signal, which is called the control signal is associated with the characteristic frequency of the controlled systems. For instance, the $C_L$ response of pitching maneuvers. In general, the last signal has a much lower frequency than the first two (Figure \ref{fig:AFC_in}b). The final actuator input signal, which is a combination of all the three signals is shown in Figure \ref{fig:AFC_in}c. It is very important to point out that the $400 Hz$ frequency has nothing to do with the dynamic system to be modeled, but is only needed to create a maximum $C_{\mu}$ from the actuators. Thus, only the lift response to the latter two signals will be modeled. 

\FloatBarrier

\section{Modeling methods}\label{subsec:result}

\begin{figure}[h]
	\centering
    \includegraphics[width=0.5\textwidth]{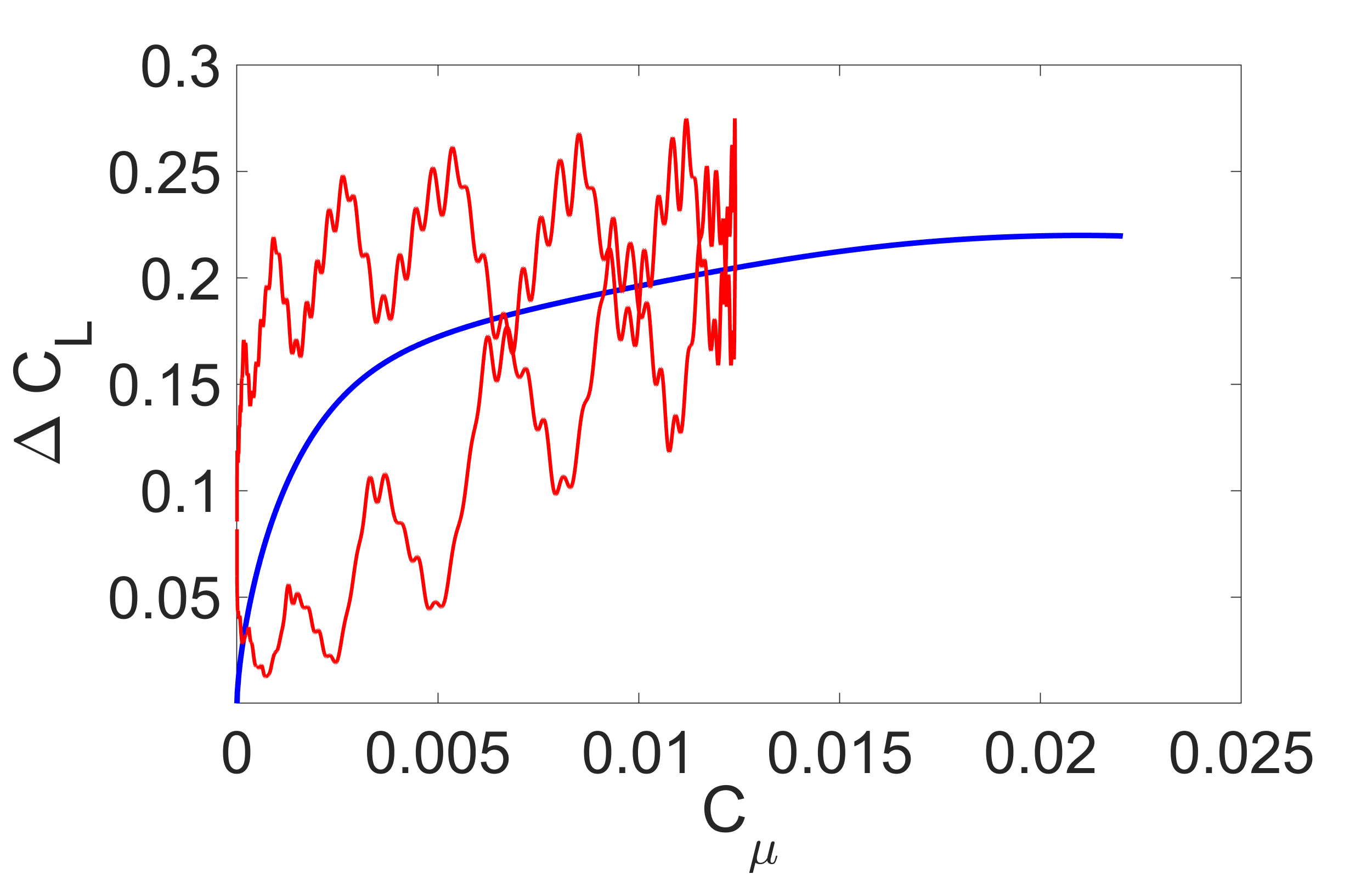}
    \caption{Lift response to static (blue) and dynamic (red) actuation.} 
    \label{fig:static_map} 
\end{figure}

Prior to modelling the lift variation, the preliminary results are examined.    
The static map is shown in Figure \ref{fig:static_map}, where the actuation $C_{\mu}$ is slowly varied from $0$ to $0.022$ and then, slowly back to $0$. The lift coefficient increment, with the non-actuated baseline subtracted, $\Delta C_L$ goes from $0$ to $0.22$ and then, back to 0 respectively. It can be seen that there is no static hysteresis loop, in other words, the static map is independent of the initial condition. However, when a fast time-varying ($k=0.125$) actuation signal is given to the actuators, the $\Delta C_L$ curve deviates from the static map, which means that the quasi-steady approach becomes inaccurate and a dynamic model is required.
\FloatBarrier

From the $\Delta C_L$ response to the dynamic actuation (the red line in Figure \ref{fig:static_map}), it is clear that there exist two characteristic features corresponding to two different frequencies. As previously mentioned in Section \ref{subsec:intro}, we will introduce a hybrid model that contains a modified convolution model for the high-frequency component of $\Delta C_L$ response to the dynamic actuation and a Wiener model for the low-frequency component.

\subsection{Modified convolution model}\label{subsec:7conv} 
\begin{figure}[ht]
	\centering
	\mbox{\subfigure[]{\epsfig{figure=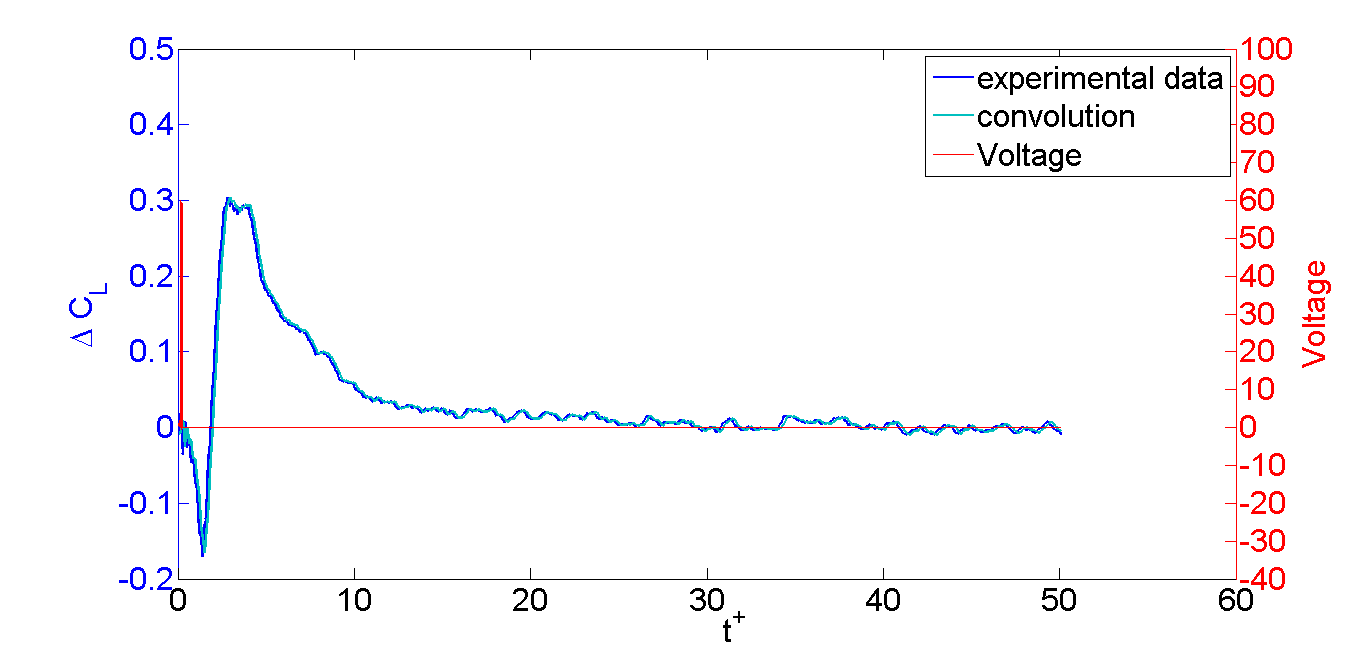, width=2.6in}}\label{fig:single}\quad
	    	\subfigure[]{\epsfig{figure=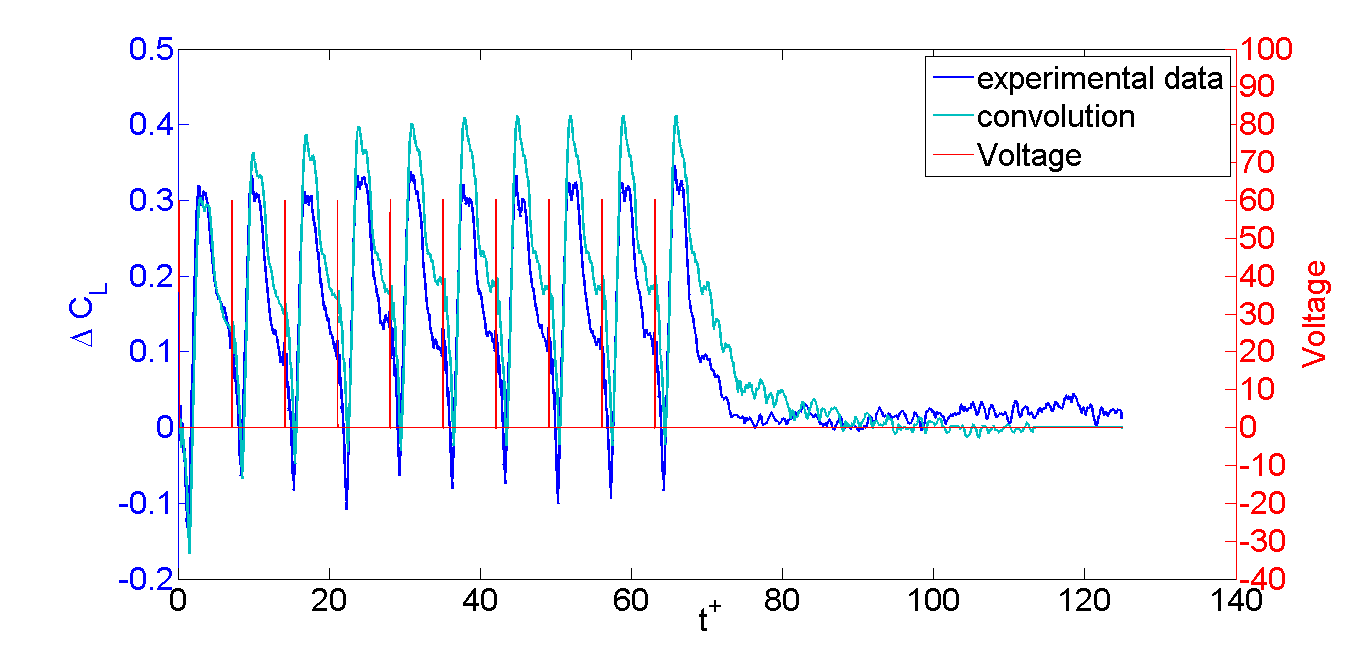,width=2.6in}}
} 
  	\mbox{\subfigure[]{\epsfig{figure=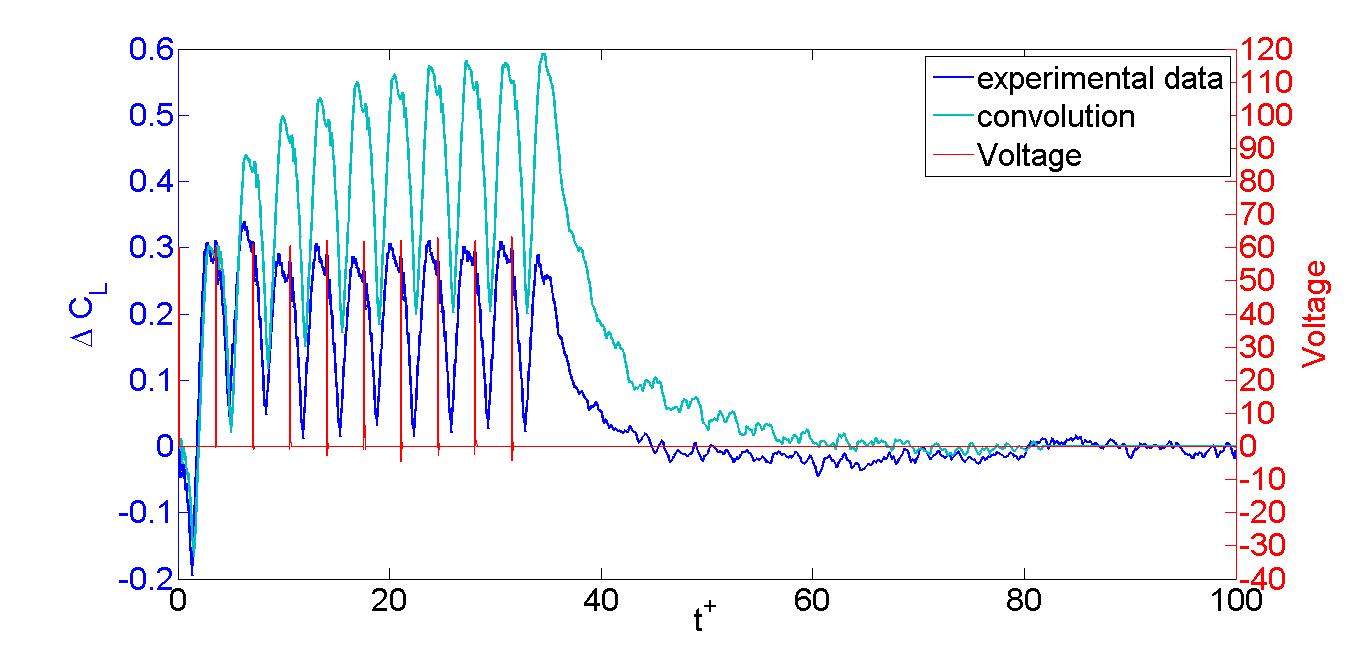,width=2.6in}}\quad
  			\subfigure[]{\epsfig{figure=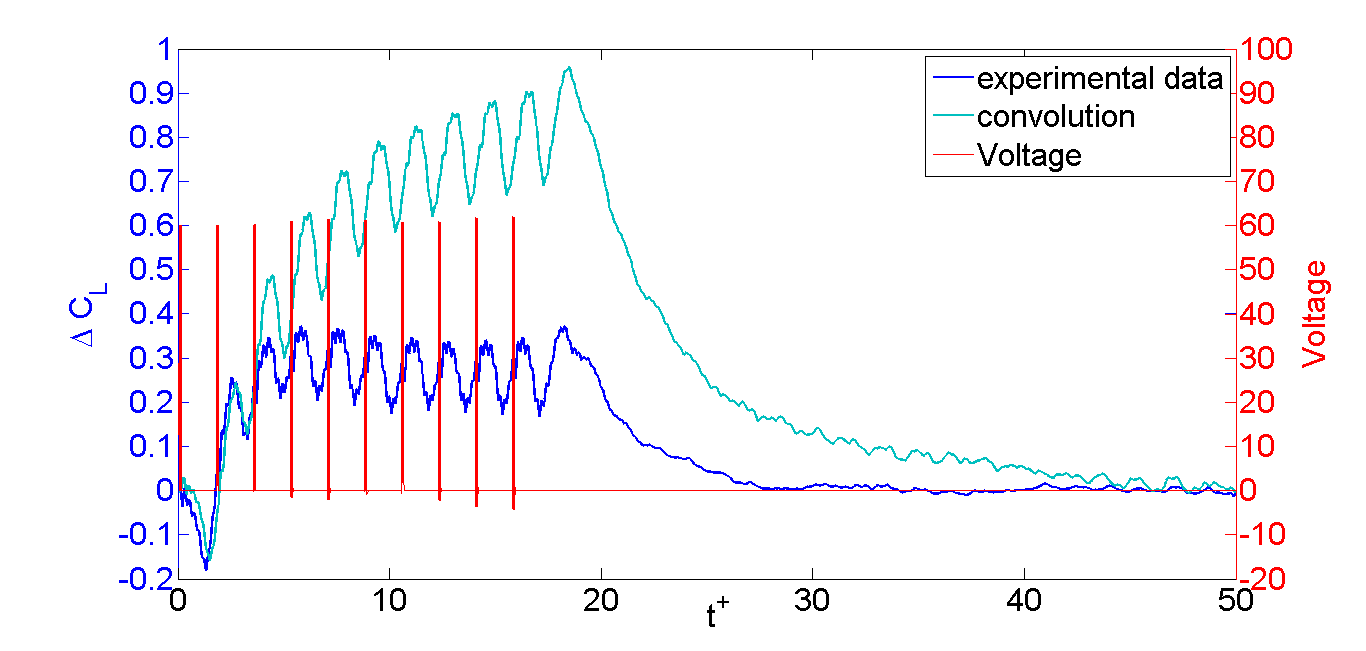,width=2.6in}}\label{fig:multi_1p75}
    	
}
	\caption{Lift response to burst-actuation at $\alpha=20^o$. The red lines show the voltage to the actuators, the blue line is the measured $\Delta C_L$ and the green line is the $\Delta C_L$ predicted by the convolution model, a) single burst at t=0; b)10 repeated bursts at the optimal $T = 1.75_{tconv}$; c) 10 repeating bursts at $T = 3.5t_{conv}$; d) 10 repeating bursts at $T = 7t_{conv}$.}
    \label{fig:multi}		
\end{figure}

The actuation signal in the current research can be viewed as an amplitude modulated burst signal, which can be seen in Figure \ref{fig:AFC_in}c. The convolution approach is based on the assumption that the burst-burst interaction is linear. Thus, the $\Delta C_L$ response to a single burst actuation can be used as the kernel of convolution as such, the convolution of the $\Delta C_L$ response to a single burst response with the discrete input signal is given by \cite{williams2010unsteady}. The convolution kernel $\Delta C_{L_{single}}$ is directly obtained from the experimental data by commending a single burst with an amplitude of $C_{\mu}=0.0022$ to the actuator.


\begin{equation} \label{eq:71}
\Delta C_L(k)=\phi\sum \Delta C_{L_{single}}(j)C_{\mu}(k-j)
\end{equation}
where $\phi$ is a constant used for normalization. The convolution model is first tested with a sequence of 10 bursts. The amplitude of the actuation is fixed at 60V corresponding to $C_{\mu}=0.0022$, $\Delta t_b=0.01s=0.125t^+$. The time interval between the bursts is varied between $1.75t^+ \leq T \leq 7.0t^+$. The convolution kernel is shown in Figure \ref{fig:multi}a, the convolution model prediction for this case is perfect, since the convolution acting on the single burst actuation just repeats itself. This also implies that when the bursts separated infinitely in time, burst-burst interactions will not arise and the convolution model is able to predict the $\Delta C_L$ variation.

However, as the bursts get closer together in time, the convolution model will to over predict the $\Delta C_L$ due to stronger nonlinear burst-burst interaction. This can be clearly seen from Figure \ref{fig:multi}b to Figure \ref{fig:multi}d. Taking a closer look at Figure \ref{fig:multi}b to Figure \ref{fig:multi}d, one can further observe that the convolution model mainly over predicts the low-frequency component (main trend) of $\Delta C_L$, but it closely tracks the high-frequency component. 

\FloatBarrier

Therefore, the convolution model is modified by adding a high-pass filter on its output, so that the modified convolution model is capable of predicting the high-frequency component only. Thus, the model for the high frequency component of the lift variation $\Delta C_{L,H}$ is given as
\begin{equation}
    \Delta C_{L,H}(k)=HF(\phi\sum \Delta C_{L_{single}}(j)C_{\mu}(k-j))
\end{equation}
where $HF$ denotes the high-pass filter. Next, we will propose a Wiener model for the low-frequency component of $\Delta C_{L,L}$. A comparison between the original convolution model and the high-pass filtered convolution model for the k = 0.

\begin{figure}[h]
	\centering
    \includegraphics[width=.6\textwidth]{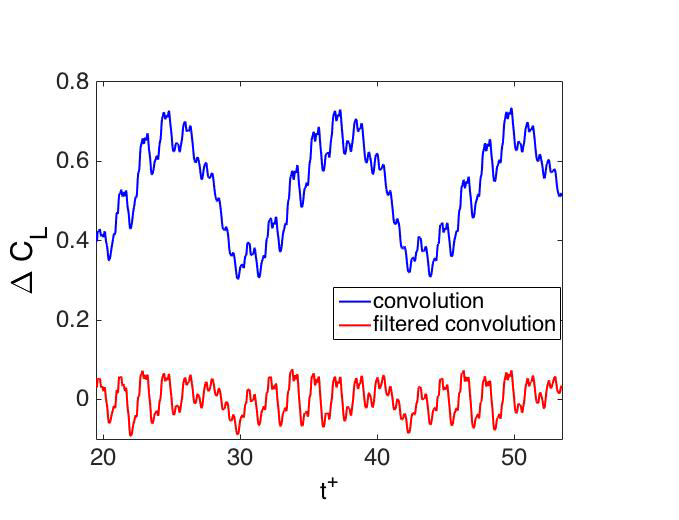}
    \caption{Standard convolution integral of the sinusoidally varying burst amplitude signal (blue line), and
    after high-pass filtering (red line).} 
    \label{fig:conv} 
\end{figure}
\FloatBarrier


\subsection{Wiener model}\label{subsec:HW} In order to model the low-frequency component of the dynamic actuation, only the low-frequency component of the actuator input signal (control signal) is used as the input to this model. Thus, the input $C_{\mu}$ in the Wiener model (which will be discussed in detail later) is only the low-frequency envelope (control signal) of the actuation signal. An et al. \cite{an2016modeling}, showed that a Goman-Khrabrov model \cite{GK} that contains a first order linear time invariant (LTI) system with a nonlinear forcing term is able to model $\Delta C_{L,L}$, the low-frequency component of $\Delta C_L$ in response to the dynamic actuation. Inspired by the Goman-Khrabrov model, we will extend this model to a more general form of a Wiener model \cite{thomson1955response}\cite{wills2013identification}, which contains a higher order LTI system with a nonlinear gain on the output. In the current research, the system identification procedure for Wiener model can be simplified from the original Wiener model. The nonlinear gain on the output can be well defined from the static experimental data shown in Fig.\ref{fig:static_map}. This nonlinear gain is defined as a lookup table $N(C_{\mu})$, where $C_{\mu}$ is the actuation momentum coefficient.
Thus, the Wiener model is expressed as 

\begin{eqnarray} 
\boldsymbol x(k+1) = \boldsymbol A \boldsymbol x(k) + \boldsymbol B C_{\mu}(k) \label{eq:LTI} \\
y(k) = \boldsymbol C \boldsymbol x(k) \label{eq:LTI1} \\
\Delta C_{L,L}(k) = N(C_{\mu}(k))y(k) \label{eq:N}
\end{eqnarray}
where $N(C_{\mu})$ is defined as 
\begin{equation} 
    N(C_{\mu})=\Delta C_{L,L}/C_\mu
\end{equation}
from the static map.

\begin{figure}[h]
	\centering
	
	\mbox{\subfigure[]{\epsfig{figure=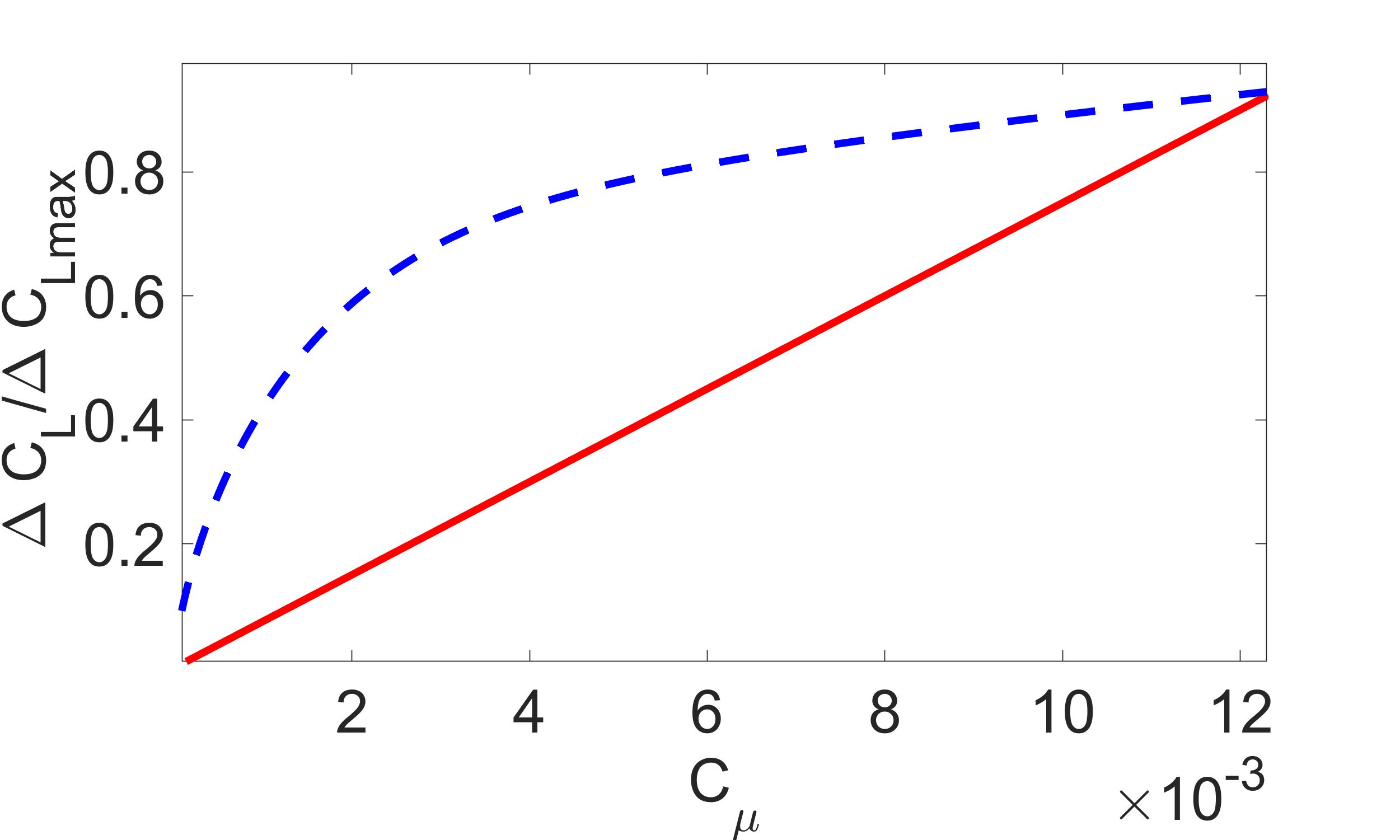, width=2.5in}}\quad} 
	~
   	\mbox{\subfigure[]{\epsfig{figure=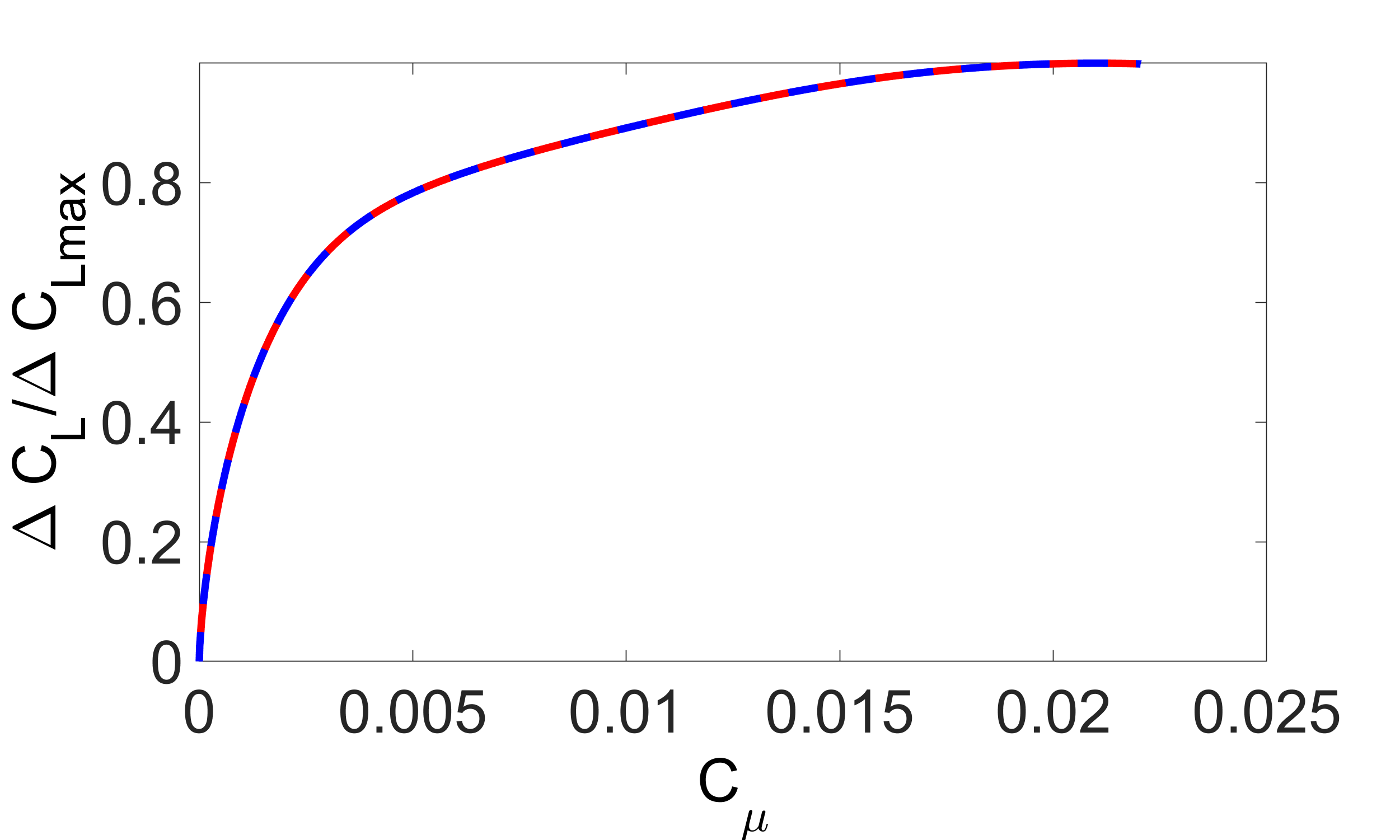,width=2.5in}}\quad}

    \caption{The static map comparison between experimental data(dashed blue line) and model(red line) (a) LTI model, (b) HW model. $\Delta C_L$ is the lift variation relative to the non-actuated lift, and $\Delta C_{Lmax}$ is defined as the maximum achievable $\Delta C_L$ associated with the maximum actuation amplitude. }
    \label{fig:static_gain}		
\end{figure}

The comparison of the static map before and after the nonlinear compensation is shown in Figure \ref{fig:static_gain}. It exhibits that the nonlinear gain, $N(C_{\mu})$ does correct the error caused by the constant gain.

\FloatBarrier

Given the nonlinear gain $N(C_{\mu})$, the LTI system (Equation. \ref{eq:LTI} and Equation. \ref{eq:LTI1}) could be identified by a variety of regression approaches. The current model is identified with the k = 0.125 sinusoidal actuation case. In the current research, we employed the prediction error method (PEM) \cite{ljung2001system} to identify the LTI model. Here, we give a brief description of the PEM, readers interested in more details about PEM please refer to \cite{ljung2001system}. 



Given the system order n, the system matrices can be parameterized by $\boldsymbol \theta$, such that Equation. \ref{eq:LTI} and Equation. \ref{eq:LTI1} can be expressed as  

  

\begin{eqnarray}
    \boldsymbol x(k+1)=\boldsymbol A(\boldsymbol \theta) \boldsymbol x(k) + \boldsymbol B(\boldsymbol \theta) \boldsymbol C_{\mu}(k) \\
    y(k) = \boldsymbol C(\theta) \boldsymbol x(k)
\end{eqnarray}
Constraining the system matrices to controllable canonical form reduces the number of parameters $\boldsymbol \theta$ to be determined.
Then, the following optimization problem can be solved to determine $\boldsymbol \theta$: 
\begin{equation}
    \min_{\boldsymbol \theta}(|y-\hat{y}|)
\end{equation}
where $\hat{y}$ is the measured output of the system. Here, $\hat{y}$ can be obtained using the $\Delta C_L$ measurement and Equation. \ref{eq:N} since $N(C_{\mu})$ is known. This minimization was then solved by the damped Gauss-Newton method that is described in \cite{ljung1999system}. A second order state-space model is obtained from the PEM. Combining the LTI system and the nonlinear gain, we can express the Wiener model as 



    

     



    \begin{align} \label{eq:SS}
     \boldsymbol x(k) &= \begin{bmatrix}
          1.9960 & -0.9960\\
          1.0000 & 0 \\
         \end{bmatrix}
         \boldsymbol x(k-1)+
         \begin{bmatrix}
          1 \\
          0 \\
         \end{bmatrix}
         \boldsymbol C_{\mu}(k)
  \end{align}
  
      \begin{align} \label{eq:y}
    y(k) &= \begin{bmatrix}
          0.0003288 & -0.0003276 \\
         \end{bmatrix}
         \boldsymbol x(k)
  \end{align}
  
  \begin{equation}
       \Delta C_{L,L}(k) = N(C_{\mu}(k))y(k)
  \end{equation}
To simplify notation moving forward, we will refer to the Wiener model as 

\begin{equation}
    \Delta C_{L,L}(k)=W(\Delta C_{L,L}(k-1), C_{\mu}(k))
\end{equation}

\begin{figure}[h]
	\centering
	
	\mbox{\subfigure[]{\epsfig{figure=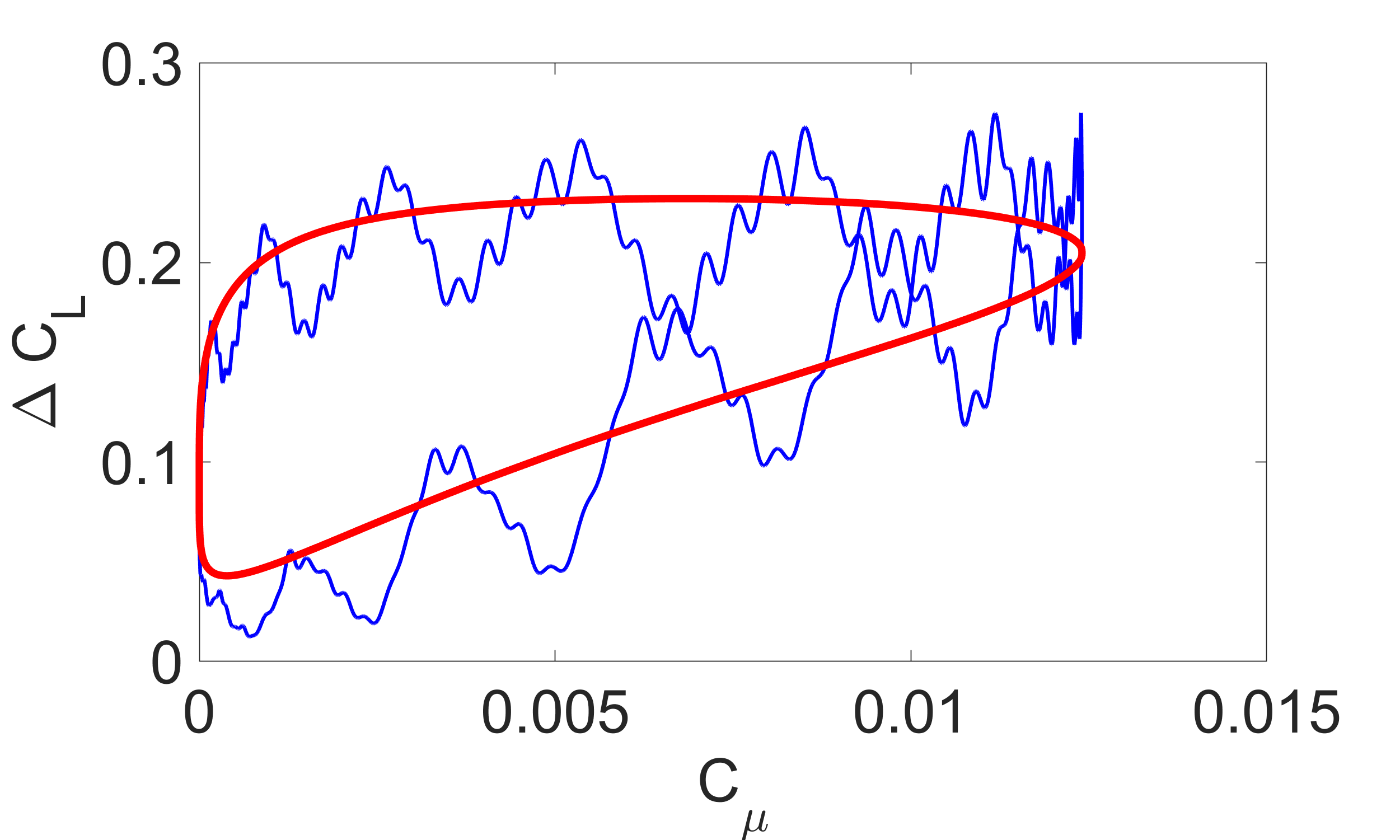, width=2.5in}}\quad} 
	~
   	\mbox{\subfigure[]{\epsfig{figure=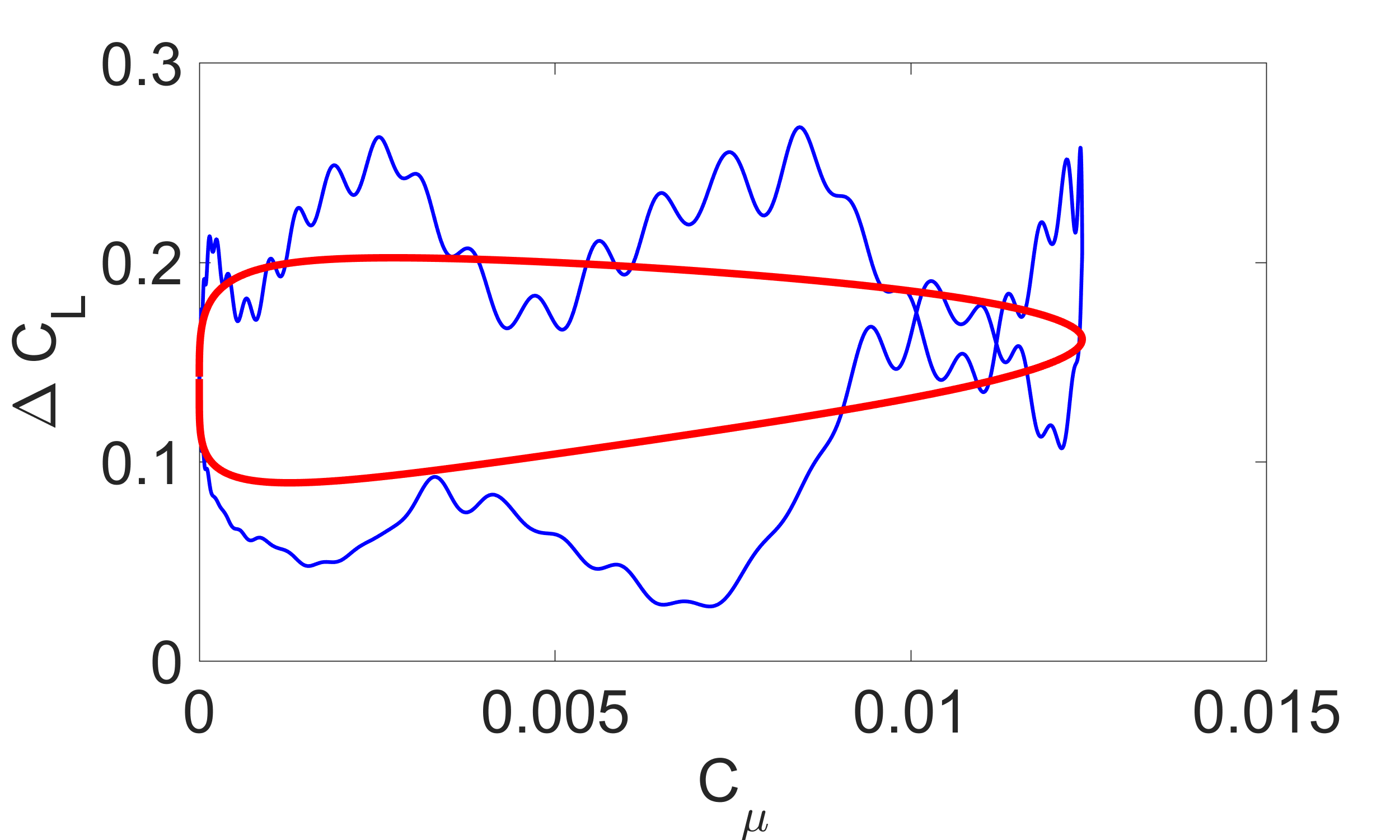,width=2.5in}}\quad}
   	
   	\mbox{\subfigure[]{\epsfig{figure=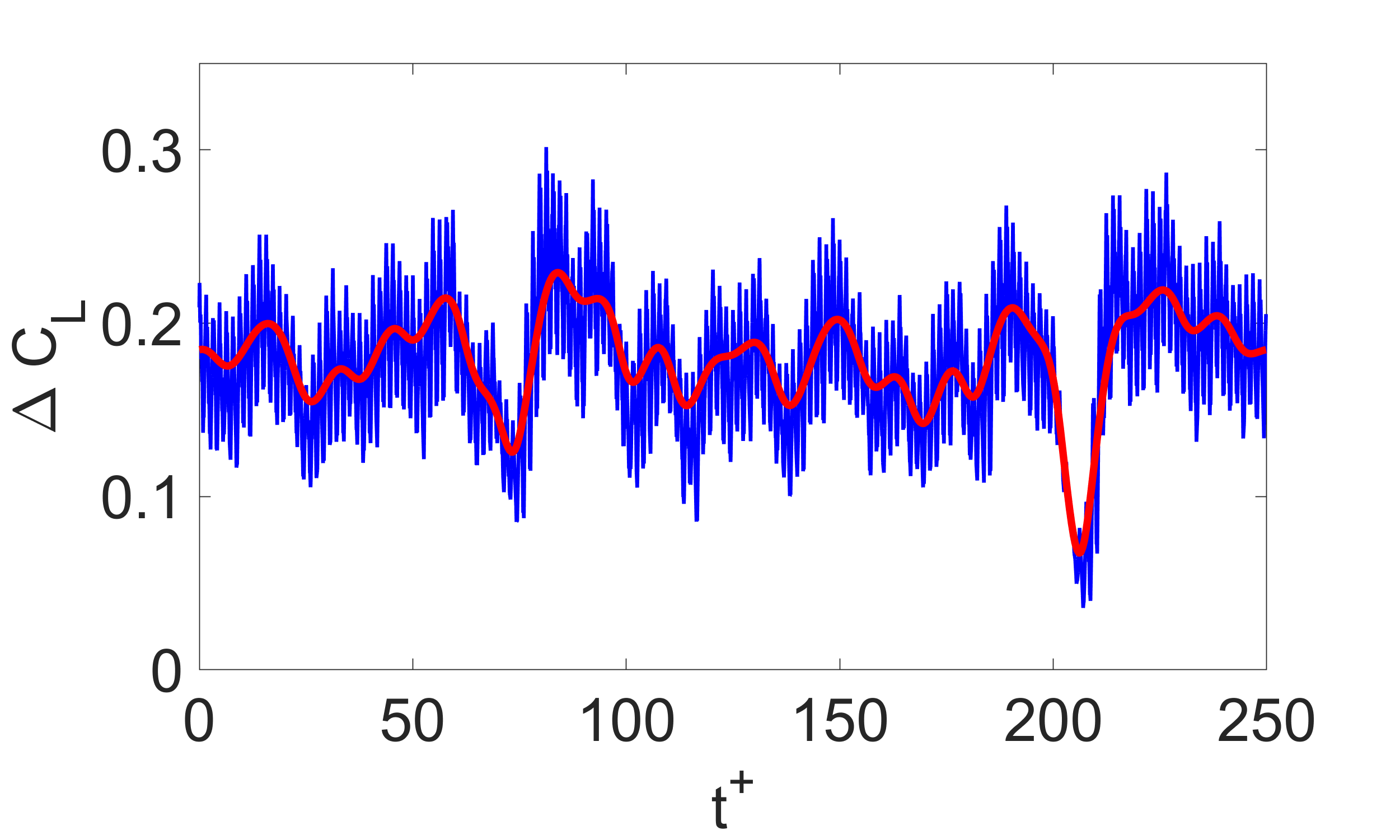,width=2.5in}}\quad}

    \caption{The comparison between the Wiener model(red line) and experimental data(blue line), (a) k=0.128, (b) k=0,25 and (c) random.}
    \label{fig:HW}		
\end{figure}

The comparisons between the Wiener model and the experimental data for both periodic and random actuation are shown in Figure \ref{fig:HW}a, Figure \ref{fig:HW}b and Figure \ref{fig:HW}c. Since the Wiener model only captures the low frequency component of the lift variation, it is hard to quantify its agreement with the experimental data. However, from Figure \ref{fig:HW}, The Wiener model is tracking the main trend of the experimental data. 

\FloatBarrier

\subsection{Hybrid Wiener-convolution model} \label{subsec:7Hybrid} As it was discussed in section \ref{subsec:7conv}, because the burst period is too short for the linear approach to be accurate, the convolution model over-predicts the amplitude of the low-frequency component, as shown by the blue line in Figure \ref{fig:multi}. But the high-frequency oscillations associated with the bursts are captured and their amplitudes are nearly correct. Thus, the signal is high-pass filtered to remove the low-frequency component.

The high-pass filtered signal is then superposed with the low-frequency component from the Wiener model. Therefore, the hybrid model that consists of the high-pass filtered convolution model and Wiener model is 
  
\begin{equation}\label{eq:hybrid}
    \Delta C_L=HF(\phi\sum \Delta C_{L_{single}}(j)C_{\mu}(k-j))+W(\Delta C_{L,L}(k-1), C_{\mu}(k))
\end{equation}
  
\begin{figure}[h]
	\centering
    \includegraphics[width=.6\textwidth]{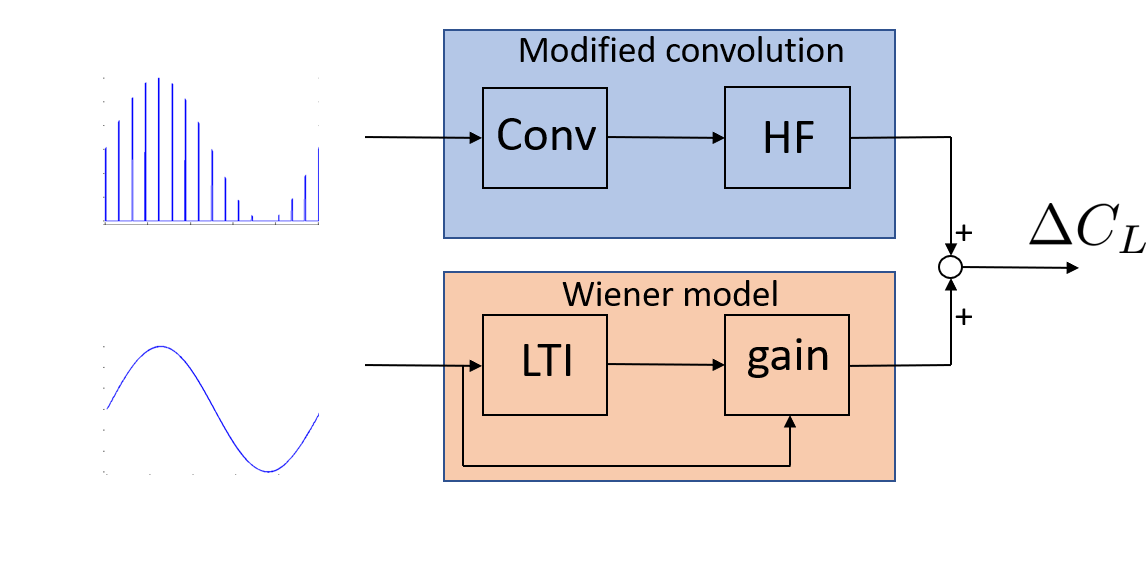}
    \caption{A flowchart of the Wiener-convolution model.} 
    \label{fig:hyb} 
\end{figure}

The first term on the right-hand side of Equation. \ref{eq:hybrid} is the high-pass filtered convolution model and the second term is the Wiener model. To better visualize the structure of this hybrid Wiener-convolution model, a flowchart is shown in Figure \ref{fig:hyb}

\FloatBarrier

\section{Model validation}\label{subsec:Results}
\begin{figure}[t!]
	\centering
	
	\mbox{\subfigure[]{\epsfig{figure=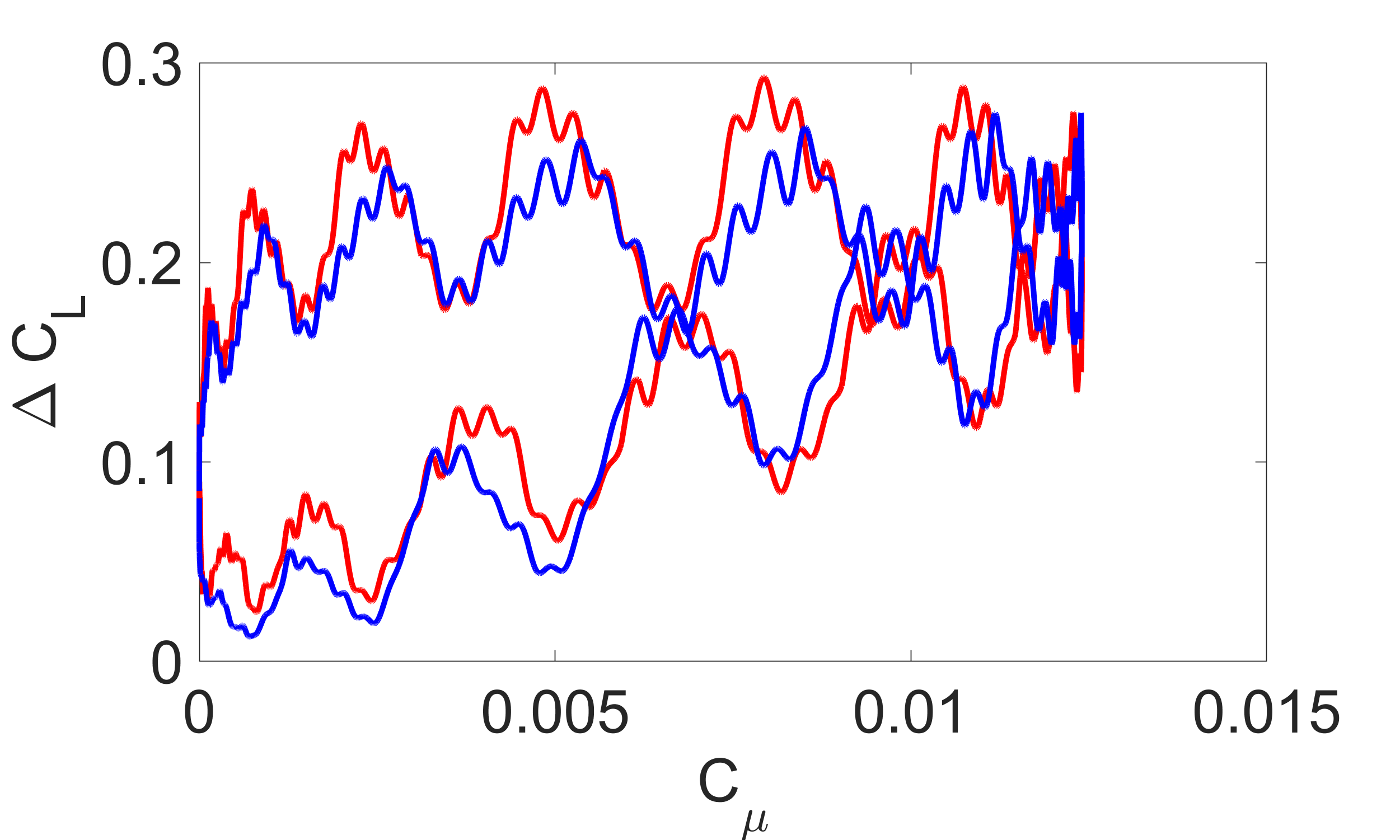, width=2.5in}}\quad} 
	~
   	\mbox{\subfigure[]{\epsfig{figure=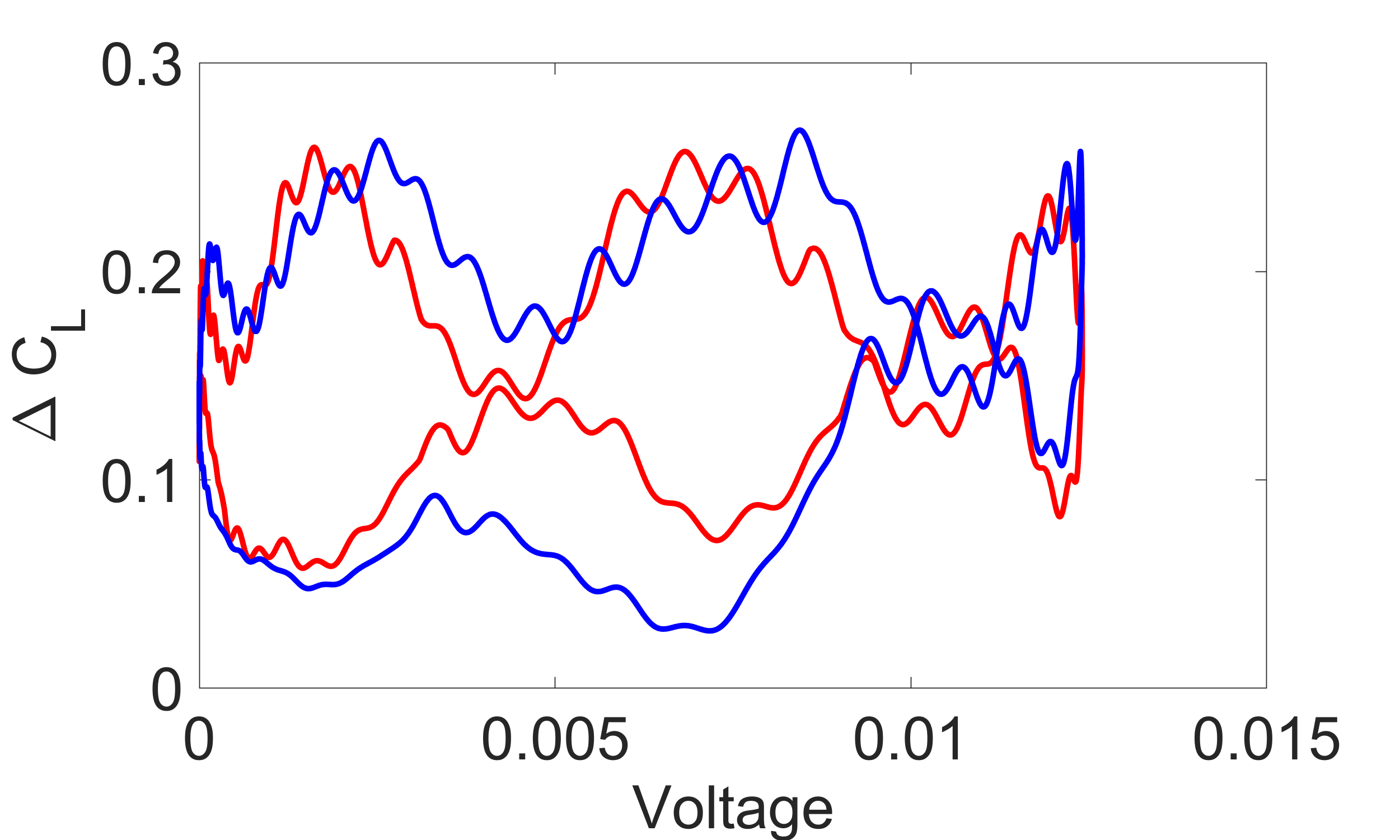,width=2.5in}}\quad}
   	
   	\mbox{\subfigure[]{\epsfig{figure=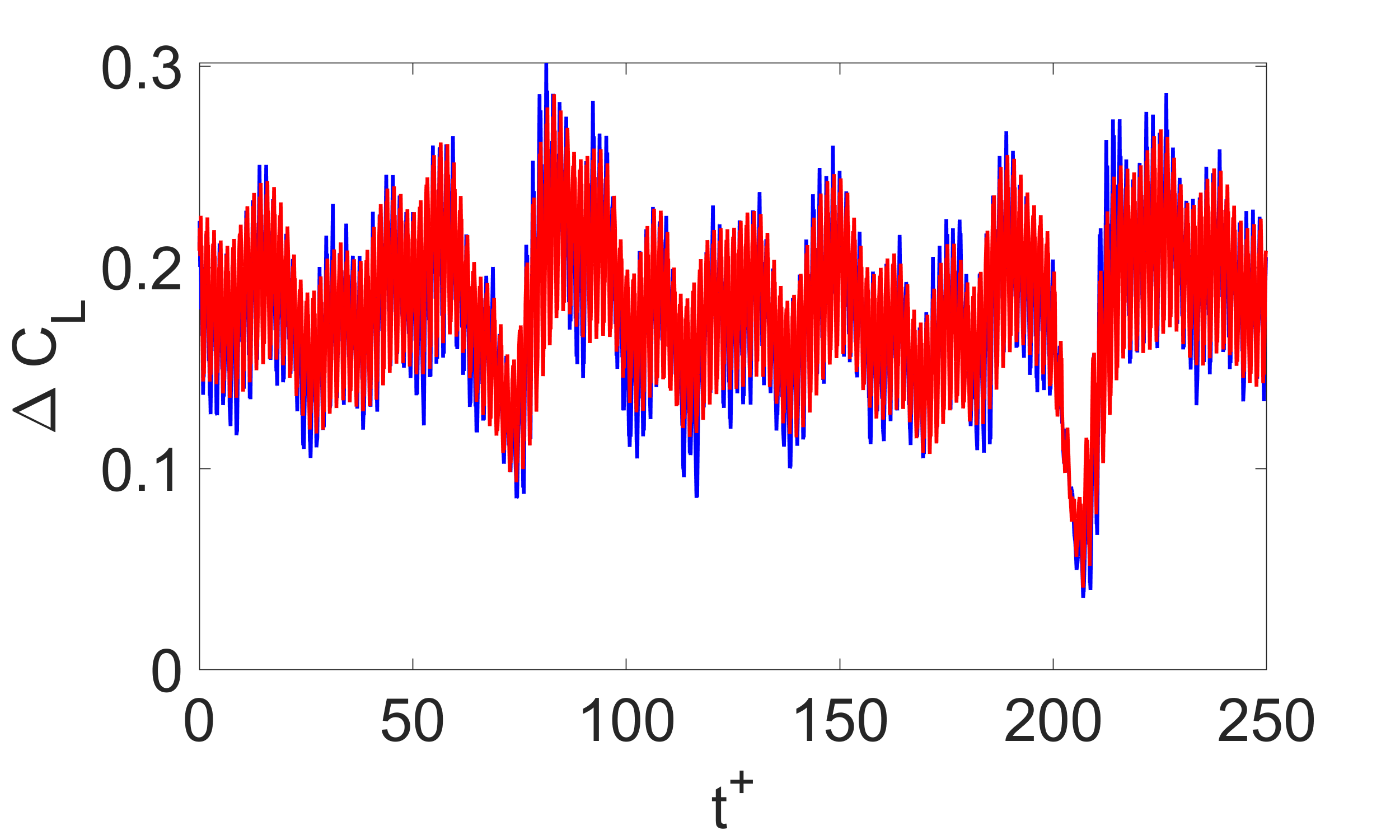,width=2.5in}}\quad}

    \caption{The comparison between the hybrid model and experimental data, (a) k=0.128, (b) k=0,25 and (c) random.}
    \label{fig:HW_Conv}		
\end{figure}

The results of the hybrid model are shown in Figure \ref{fig:HW_Conv} for two different forcing
frequencies, k = 0.128 and k = 0.25 and for a random actuation. The model was identified on the k = 0.128 case. The experimental data is shown by the blue lines, which are compared to the combined model (red lines). The combined model captures both the high frequency and low-frequency lift components. To quantitatively evaluate the hybrid model, the correlation coefficients are computed for each individual case. If the correlation coefficient between two signals is 0, then the two signals are not related. If the correlation coefficient is 1, then the two signals are completely linear dependent. Assuming we have two signals A and B, the correlation coefficient is defined as 

\begin{equation}
    \rho(A,B)=\frac{cov(A,B)}{\sigma_A\sigma_B}
\end{equation}

where $\rho(A,B)$ is the correlation between signal $A$ and $B$, $cov(A,B)$ is the covariance of $A$ and $B$, and $\sigma_A$ and $\sigma_B$ are standard deviation of $A$ and $B$. The resulting correlation coefficient between the model and measurement is 0.9596 for the case k = 0.128, 0.8654 for k = 0.25 and 0.9277 for random actuation. It is clear that the hybrid model is able to predict the lift variation due to the dyanmic actuation.

\FloatBarrier

\section{Conclusion}\label{subsec:conclusion} 

In the current research, a Wiener-convolution hybrid model was introduced and examined using a NACA 0009 airfoil with flow control actuators that are operated with open-loop forcing. The `burst mode' of actuation is used to obtain large lift increments, and the time-varying amplitude of the burst signal is modulated to obtain a time-varying (dynamic) lift coefficient. 

A convolution model utilizing the lift response to the impulse input and the actuator input signal over predicts the low-frequency (main trend), but it is capable of capturing the high-frequency component. Therefore, a high-pass filter is added on the output end of the convolution model to only model the high-frequency component of the lift variation.  

A Wiener model is then identified using the prediction error method. The nonlinear static gain in this model is used to compensate for the static nonlinearity and the linear transfer function is capable of capturing the dynamic feature of the dynamic system. 

The Wiener-convolution hybrid model obtained by combining the Wiener model and the high-pass filtered convolution model was validated with the experimental data taken from the wind tunnel. The comparison between the model prediction and the experimental data exhibits the good performance of the Wiener-convolution hybrid model. And thus, a model based feedback controller or an observer can be built utilizing the Wiener-convolution hybrid model.


\vspace{6pt} 


\authorcontributions{conceptualization, X.A., D.R.W. and M.S.H.; methodology, X.A., D.R.W. and M.S.H.; software, X.A.; validation, X.A. and D.R.W.; formal analysis, X.A. and D.R.W.; investigation, X.A. D.R.W.; resources, D.R.W.; data curation, X.A. and D.R.W.; writing--original draft preparation, X.A.; writing--review and editing, X.A., D.R.W. and M.S.H.; visualization, X.A.; supervision, D.R.W.; project administration, D.R.W.; funding acquisition, D.R.W}

\funding{This research was funded by the Air Force Office of Scientific Research under grant number FA9550-18-1-0440.}

\acknowledgments{The first author would like to thank the Office of Naval Research for additional support under grant number N00014-14-1-0533. MSH acknowledges support from the Air Force Office of Scientific Research under award number FA9550-19-1-0034.}

\conflictsofinterest{The authors declare no conflict of interest.} 

\bibliographystyle{unsrt}
\bibliography{mybib}

\end{document}